\documentclass[conference]{IEEEtran}
\usepackage{macros}

\usepackage{url}

\usepackage{breakurl}
\usepackage[breaklinks]{hyperref}

\usepackage{cite}
\usepackage{amsmath,amssymb,amsfonts}
\usepackage{graphicx}
\usepackage{textcomp}
\usepackage{xcolor}
\def\BibTeX{{\rm B\kern-.05em{\sc i\kern-.025em b}\kern-.08em
    T\kern-.1667em\lower.7ex\hbox{E}\kern-.125emX}}
\begin{document}

\title{Automated Testing of Broken Authentication Vulnerabilities in Web APIs with \securest}
% \thanks{Identify applicable funding agency here. If none, delete this.}

\author{\IEEEauthorblockN{Davide Corradini
% \IEEEauthorrefmark{1} \\ does it matter?
}
\IEEEauthorblockA{
% \textit{Computer Science Department} \\
\textit{University of Luxembourg}\\davide.corradini@uni.lu
}
%\and
%\IEEEauthorblockN{Mohammad Danish}
%\IEEEauthorblockA{
% \textit{dept. name of organization (of Aff.)} \\
%\textit{Technische Universität Clausthal}\\
%Germany\\
%mohammad.danish@tu-clausthal.de}
\and
\IEEEauthorblockN{Mariano Ceccato}
\IEEEauthorblockA{
% \textit{Computer Science Department} \\
\textit{University of Verona}\\
mariano.ceccato@univr.it}
\and
% \IEEEauthorblockN{4\textsuperscript{th} Given Name Surname}
% \IEEEauthorblockA{\textit{dept. name of organization (of Aff.)} \\
% \textit{name of organization (of Aff.)}\\
% City, Country \\
% email address or ORCID}
% \and
\IEEEauthorblockN{Mohammad Ghafari}
\IEEEauthorblockA{
% \textit{dept. name of organization (of Aff.)} \\
\textit{Technische Universität Clausthal}\\
mohammad.ghafari@tu-clausthal.de}
% \and
% \IEEEauthorblockN{6\textsuperscript{th} Given Name Surname}
% \IEEEauthorblockA{\textit{dept. name of organization (of Aff.)} \\
% \textit{name of organization (of Aff.)}\\
% City, Country \\
% email address or ORCID}
}

\maketitle

\begin{abstract}
% Web APIs have become integral components of modern software systems, facilitating communication and data exchange between various applications. Despite their widespread use, web APIs often contain defects and vulnerabilities, highlighting the need for effective testing methodologies.

We present \securest, an open-source security testing tool targeting broken authentication, one of the most prevalent API security risks in the wild. \securest automatically tests web APIs for credential stuffing, password brute forcing, and unchecked token authenticity.
Empirical results show that \securest is effective in improving web API security. Notably, it uncovered previously unknown authentication vulnerabilities in four public APIs.

\noindent
Demo video: \url{https://tiny.cc/authrest-video}

\noindent
Source code: \url{https://github.com/SeUniVr/AuthREST}
\end{abstract}


\section{Introduction}
Web APIs, such as REST APIs, have become indispensable components in modern web architectures, serving as gateways for remote clients to integrate functionality and access resources over the Internet.
Web APIs drive a significant portion of internet traffic and enable critical services, making them an attractive target for attackers.
The latest Salt Security report shows that web API attacks have increased by 400\% in just a few months~\cite{saltsecurityreport}. The report also reveals that 99\% of respondents face challenges in managing API-related incidents, and 22\% of organizations experienced a data breach. 
These findings are consistent with prior research, which highlights weak security practices in API development, on both the client~\cite{Gadient2020} and the server~\cite{Gadient2021} sides. 

%
%This gap has led to serious consequences in recent years.
%Tens of millions of T-Mobile customers' data--including payment card information, social security numbers, and government ID numbers--were compromised via an API exploit~\cite{tmobile}.
%
%Hackers exploited a vulnerability in the X (formerly Twitter) API, and exposed information of 5.4 million user accounts. The vulnerability originated from an API designed to help users find others, which enabled hackers to submit email addresses or phone numbers to the API and retrieve the associated X account~\cite{attack5milion}.
%
%The personal details of 22.5 million Malaysians were on sale due to unauthorized access to myIDENTITI, the API  that provides government agencies with information about Malaysian citizens~\cite{attack1milion}.
%
%FlexBooker reported an API breach in its AWS S3 bucket that leaked 3.7 million user records~\cite{attack3milion}. 
%
%Finally, 
%with the Optus API beach, attackers discovered a publicly exposed endpoint that did not require authentication~\cite{attack2milion}. 
%It is estimated that this breach exposed 11.2 million customer records, and it incurred \$140 million loss.

Automated API security testing has emerged as a promising activity to enhance API security and streamline the development process for API developers. 
However, only a few approaches have been proposed in the literature to automatically test specific vulnerabilities of web APIs, for instance, mass assignment~\cite{corradini2022massassignment,Mazidi2024}, excessive data exposure~\cite{edefuzz}, and SQL or command injection~\cite{deng2023nautilus,duvulnerability}.
Unfortunately, there are no completely automated test case generation techniques for the (many) remaining well-known vulnerabilities in web APIs.
According to the Salt Security report~\cite{saltsecurityreport}, the most frequently identified issues over the past year include password brute forcing, credential stuffing, and other authentication-related problems.
OWASP Foundation has also listed \textit{broken authentication} as one of the major API security concerns~\cite{owasp2023brokenauth}.

To fill in this gap, we introduce \securest, an open-source tool that implements three testing strategies for broken authentication, namely credential stuffing, password brute forcing, and unchecked token authenticity. 
\securest adopts a black-box approach, \ie it only requires access to the HTTP interface of the API and its specification, making it applicable to any web API independently of the programming language and framework used for its implementation.
We conducted experiments with \securest and uncovered four previously unknown authentication vulnerabilities in four publicly accessible web APIs, showing its potential to enhance API security in real-world settings.

\section{\securest}
\label{sec:framework}

\begin{figure}[t]
    \centering
\includegraphics[width=0.5\textwidth]{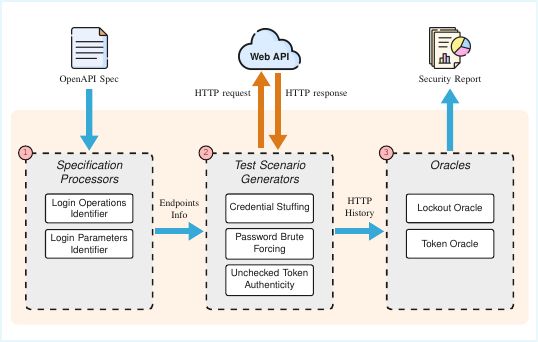} 
    	\caption{The architecture of \securest.}\label{fig:approach-overview}
\end{figure}

\securest is a security testing tool designed to identify vulnerabilities related to credential stuffing, password brute-forcing, and the use of unauthorized tokens in web APIs. To the best of our knowledge, it is the only tool that automatically generates comprehensive test cases for broken authentication, identifying vulnerable API operations and parameters without human intervention.

\subsection{Architecture}
Figure~\ref{fig:approach-overview} illustrates the architecture of \securest, built on top of the RestTestGen framework~\cite{corradini2022resttestgen}.
It takes as input the \openapi specification of the API under test, allowing it to extract or infer relevant information for generating test cases. It then creates black-box security test cases, which consist of sequences of HTTP interactions with the API. Finally, the HTTP interaction history is evaluated by oracles, producing a security report that details any identified vulnerabilities.

% \securest implements various specification processors, test scenario generators, and oracles. 
% We introduce these components for each security testing scenario.

\subsection{Credential Stuffing Security Testing}
The primary goal of this testing strategy is to automatically determine whether the API under test is vulnerable to credential stuffing attacks. If the login endpoint (also known as ``login operation") lacks effective lockout mechanisms, an attacker can repeatedly attempt logins using a predefined set of credentials. In contrast, a secure API detects excessive requests from the same client and responds by imposing rate limiting, blocking, challenge mechanisms, or requiring two-factor authentication before permitting further login attempts.

\subsubsection{Specification Processing}
% Credential stuffing targets login operations, specifically the user identifier and password parameters. 
To generate effective test scenarios, it is essential to identify which operations and parameters are relevant for user authentication.
To this end, we reviewed the specifications and documentation for 100 APIs from APIs.guru and, for those that supported authentication, we collected the typical syntax used for login operations.
The outcome enabled us to develop two heuristics, which we explain in the following.

\paragraph*{Login Operations}~We found that a majority of login operations are invoked using the HTTP methods \smalltt{POST} and \smalltt{GET}, and the path often includes the keywords ``login'' or ``sign in'' in various forms such as ``login'' or ``log-in''. %Indeed, in the \textit{Bank API} example, the syntax of the login operation is \smalltt{POST\,/login}.
We concluded that login operations often adopt recurring patterns, and a simple heuristic to search for relevant keywords is effective in locating them.
Hence,
we classify an operation as a login operation when a \smalltt{POST} or \smalltt{GET} method is used on a \emph{path} that contains the ``login'' or ``signin'' strings, after removing non-alphabetical characters (\eg ``-'' or ``\_'' characters).

\paragraph*{Login-Related Parameter Names}~We collected the parameter names associated with login operations, specifically those instrumental in transmitting user identifiers and passwords. 
We found that they are typically defined as string type, occasionally as numeric type, but never as booleans, arrays, or objects. 
In most APIs, the parameter associated with the password was just named ``password'', and in a few instances, the term ``pass'' was used. 
However, naming conventions for user identifiers were more diverse, including terms such as ``user'', ``username'', ``email", ``mail'', ``phone'', ``login'', ``userId'', or simply ``id''.

In response to this variability, we developed a heuristic based on priority. 
For passwords, we prioritize parameters whose names contain the full string ``password'', but if none match, we consider any other parameter whose name contains ``pass''. 
For user identifiers, we prioritize parameters with names containing (not matching!), in the following order, ``username'', ``email'', ``login'', ``user'', ``phone'', ``mail'', and finally ``id'', after having removed non-alphanumerical characters from parameter names. Notably, a parameter called ``userId'' would first match the substring ``user'', fourth in the priority scale, and not ``id'', last in the scale. This allows the heuristic to correctly guess the user identifier among other resource identifiers, possibly used in the operation.

\subsubsection{Test Scenario} To simulate a credential stuffing attack, the test scenario involves generating a sequence of HTTP requests directed towards the previously identified login operation. All parameters associated with this endpoint must be appropriately assigned a value. We assign random values to the user and password parameters when they are automatically identified. It is important to note that we are not conducting actual credential stuffing, as we generate credentials from scratch. Nevertheless, from the API's perspective, this is indistinguishable from an actual attack.

It is worth mentioning that a login endpoint may accept additional parameters beyond user identifiers and passwords. To supply valid values for these parameters, we leverage the state-of-the-art parameter input value providers available in \rtg Framework~\cite{corradini2022resttestgen}. 
For example, we use values directly from the \openapi specification (\eg documented example values, enum values), values observed in previous successful HTTP interactions with the API (from both requests and responses), and random values.

We send 100 requests to the API within 10 seconds. This high volume of requests within a short time frame typically indicates malicious activity and should activate mitigation countermeasures.
In fact, according to best practices, countermeasures should be taken after receiving a rapid sequence of 3-10 requests from the same source~\cite{owasp2023brokenauth}.

\subsubsection{Oracle} The oracle's role involves analyzing the HTTP execution trace and reporting a security defect when the API fails to implement a lockout mechanism. To declare a failure, we assess two critical properties, recommended as best practices by the OWASP Foundation~\cite{owasp2023brokenauth}:

\begin{enumerate}

    \item The API does not force rate limiting, as indicated by the absence of a \fourtwonine status code in the responses.

    \item  Despite a series of login attempts, the error messages remain the same, whereas they should change when effective lockout mechanisms are implemented.
    For instance, a basic ``Wrong username or password'' error message might be altered to something like ``Please try again later'' due to such mechanisms being active.
    
\end{enumerate}

While verifying the first property is relatively straightforward, assessing the second property poses greater complexity. This is because even if error messages are semantically identical, they may appear different when compared at the string level. 
% This discrepancy can arise from the inclusion of timestamps, unique error identifiers, IP addresses, or other metadata, which may inadvertently mislead the similarity-check algorithm into categorizing two error messages, which are indeed semantically identical, as distinct. 
This discrepancy can result from timestamps, unique error IDs, IP addresses, or other metadata that cause the similarity-check algorithm to mistakenly treat semantically identical error messages as different.
Therefore, we adopt Porter's stemming algorithm to tokenize error messages~\cite{porter1980algorithm}, and we consider two error messages to be related to the same error if they share at least 70\% of the tokens.
We adopted this threshold based on the conservative consideration that if two messages are quite (\ie at least 30\%) different, then they are about different errors. The remaining similarity may be due to similar context, \eg both errors could be about the authentication on the same web API.

\subsection{Password Brute Force Security Testing}
This testing strategy evaluates whether a web API is vulnerable to password brute force attacks during login. The attack, countermeasures, and testing approach closely resemble those for credential stuffing. We generate a sequence of HTTP requests targeting the login operation, constructed similarly to those for credential stuffing. The key difference is that the user identifier parameter remains constant across all requests, while the password parameter is randomly generated for each attempt.

% This testing strategy aims to assess whether a web API is vulnerable to password brute force attacks during the login operation. The attack, countermeasures, and, consequently, the testing strategy implementation closely resemble what we previously introduced for credential stuffing. However, there is a small but crucial difference in the test scenario.
% %
% To simulate a password brute force attack, we still generate a sequence of HTTP requests directed to the login operation, and this sequence is constructed similarly to those related to credential stuffing. However, the user identifier parameter remains the same across all requests, while the password parameter is randomly generated for each attempt. 

\subsection{Unchecked Token Authenticity Security Testing}
Web API authentication may rely on an authentication token attached to requests instead of a username-password pair
The objective of this strategy is to assess the web API's ability to effectively conduct appropriate checks on incoming tokens.
If web APIs accept invalid tokens as valid, unauthorized access to confidential resources may occur, compromising data integrity or restricted functionalities. Secure APIs should deny such requests by returning status codes \fouroone or \fourothree.

\subsubsection{Specification Processing} 
We need to explore the API under test exhaustively. To achieve this, we parse the API's specification to collect information about the available operations and their input parameters.

\subsubsection{Test Scenario} 
Initially, we execute a standard nominal testing session to explore the web API and gather information about its nominal behavior.
We adopt the nominal testing strategy proposed in the state of the art~\cite{Corradini2022}, which orders the execution of API operations according to data dependencies (producer-consumer relations among operations) and uses input data from various sources such as the \openapi specification (\eg example values, enum values, default values, etc.), past HTTP interactions, and random generators. 
%
%\mg{please read}
Nominal testing is fundamental to this strategy, as it enables deep exploration of the API by generating valid sequences of HTTP requests. It helps identify the appropriate input data needed to construct successful request sequences for later use. Throughout the nominal testing session, we use the correct token provided by the user and record the resulting HTTP interactions for future reference.
% Nominal testing is fundamental to this strategy because it enables a deep exploration of the API by generating valid sequences of HTTP requests. This process helps determine the appropriate input data needed to craft successful sequences requests for later. Throughout the nominal testing session, we use the correct token provided by the user.  Additionally, we record the sequence of HTTP interactions generated during the session for future reference.

% Throughout this nominal testing session, we use the correct token as provided by the user. Nominal testing is fundamental to this strategy because it enables a deep exploration of the API, promoting the generation of valid HTTP requests that can be reused later. 
% For example, it helps determine what input data should be used to craft a successful request. 
% We also record the sequence of HTTP interactions generated during this session for future reference.

Subsequently, we replay the recorded sequence, mutating the correct token with various operators so that each HTTP interaction uses a different mutation.
We apply the following mutations:

\begin{itemize}
    \item Altering a single character of the valid token, ensuring not to alter structural tokens (like dots) in JWT tokens, and using only valid characters (\eg not using special symbols).
    \item Removing one random character from the valid token, ensuring not to remove structural elements like dots in JWT tokens.
    \item Adding one random (valid) character to the valid token.
\end{itemize}

This sequence with mutated HTTP interactions provides the oracle with the means to observe how the API behaves when presented with incorrect (i.e., mutated) tokens.

However, API operations that do not require authentication typically ignore both valid and invalid tokens in incoming requests and are processed successfully with status code \twoxx.
Hence, it is possible that our strategy flags an operation as vulnerable solely because the operation does not mandate authentication, resulting in a reported vulnerability that is a false positive. 
To differentiate genuine vulnerabilities from false alarms, we replicate the same HTTP sequence for the third time, omitting the token from requests. We expect that only operations not requiring authentication will be accepted and successfully executed by the web API in this case. This provides us with a list of the API operations that do not require authentication and that we can safely remove from the list of vulnerable operations revealed by the approach.
% \mg{should we mention that operations that are (mistakenly) not  protected cannot be discovered .. ?}
% \davide{this is right, however we are looking for operations that do not check the authenticity of tokens, not those that do no implement auth at all by mistake}

\subsubsection{Oracle} The oracle's role is to analyze the HTTP execution trace of interactions with the web API and to report a failure when the web API deviates from its intended behavior. This includes instances where the web API processes requests containing mutated (i.e., incorrect) tokens. To evaluate the API's behavior, the Oracle computes two sets of operations from two of the execution traces previously recorded.

The set $W$ includes operations that were successfully executed (status code \twoxx) despite providing the \textit{incorrect} token in the requests. This set may include operations that do not require authentication because they typically ignore (incorrect) tokens.
The set $N$ includes operations that were successfully executed (status code \twoxx) when no token was provided in requests, meaning that these operations do not require authentication.
The final set of vulnerable operations is therefore computed as $V = W - N$, which consists of all the operations accessed with an invalid token, excluding those that do not require authentication.

\section{Empirical Validation}\label{sec:evaluation}

We investigate how effective \securest is in uncovering broken authentication in web APIs.
We (i)~quantify the number of identified vulnerabilities and (ii)~manually examine the sequence of HTTP interactions that \securest generated along with the reported vulnerabilities to assess the accuracy of the results.

\subsection{API Benchmark}

%\mg{You mention 6 APIs and then later in the paragraph mention 50 APIs!!}
We built a benchmark comprising six publicly accessible real-world APIs. 
These APIs were sourced from the repository provided by APIs.guru, %\footnote{\url{https://apis.guru/}} 
that catalogs public web APIs along with their corresponding \openapi specifications (a prerequisite for our approach). In particular, we started by randomly sampling 50 web APIs from this catalog to manually check their suitability for our experiment (these APIs are different from those considered while building our heuristics).
%We checked that APIs were reachable by contacting the server at the URL specified in the \openapi specification.
We excluded unreachable or non-working APIs, %for example, the \textit{Trivial Life API},\footnote{\url{https://tl-api.azurewebsites.net}} which always responds with \fourofour status codes when contacted.
% 
% Also, we removed APIs
and those that do not require authentication.
 % as our objective is to test broken authentication.
In addition, we excluded APIs that do not explicitly list a login operation in the API's specification.
%
% These steps were conducted manually by inspecting the \openapi specifications, rather than in an automated manner, to mitigate any potential bias in the experiment regarding the accuracy of automated detection of the login operations and related parameters. 

\begin{comment}
\begin{table}[t]
\centering
\caption{The APIs in our benchmark.}
\label{tab:api-benchmark}
\renewcommand\arraystretch{1.1}
\begin{tabular}{l|c|c}
\textbf{API Name} & \textbf{URL} & \textbf{N. Operations}  \\ \hline\hline
Here & tracking.api.here.com & 144 \\
ID4I & backend.id4i.de & 107 \\
6 Dot & 6-dot-authentiqio.appspot.com & 14 \\
BeezUP & api.beezup.com & 195 \\
BRAINBI & brainbi.dev & 14 \\
Tradematic & api.tradematic.com & 67 \\
\end{tabular}
\vspace{-2mm}
\end{table}
\end{comment}

%\newpage
In total, we collected six public APIs suitable for this study: Here, ID4I, 6 Dot, BRAINBI, BeezUP, and Tradematic. These APIs vary in size and complexity, as measured by the number of operations and parameters, and they belong to different domains. The benchmark APIs, along with their specifications and the number of exposed operations, are available on the \securest GitHub page.

%\mg{I think it is important to also mention the ethical considerations ... }

% The size and complexity of this benchmark are consistent with the existing literature on REST API testing~\cite{10.1145/3617175}. Furthermore, we consider it to be representative of real-world APIs, as the included APIs are diverse in size and complexity (in terms of the number of operations and parameters) and belong to different domains.

\subsection{Experimental Procedure}
% The testing strategies implemented with \securest have been executed on each API in the benchmark (credential stuffing and password brute forcing strategies on five APIs and the unchecked token authenticity strategy on one API). 
We applied \securest to the APIs in the benchmark.
As the tool includes non-deterministic components such as random input value generation, we repeated experiments 10 times to control non-determinism. Consequently, the results report average values (true positives, false positives) across these ten executions.

% Results of automated identification of login operations and related parameters were stored for metrics collection and computation.

%\mg{Looks good. Isn't this more relevant to Experiment Procedure?}

%For unchecked token authenticity, we closely monitored the experiments to prevent any unwanted side effects, and in all cases, no vulnerable APIs were found.
%\mg{the last part after the comma where you mention we found nothing, may not be necessary here. However, you can mention it in the study result}

The security reports flagging a vulnerability were subject to manual inspection to validate their accuracy. Specifically, the authors of this paper analyzed the HTTP interaction history that revealed the vulnerability to confirm it. Additionally, we manually exploited the vulnerability using the third-party tool Postman %\footnote{\url{https://www.postman.com/}}
to confirm its presence.

For ethical considerations, our strategies perform only simulated credential stuffing and password brute forcing attacks, limiting our tests to checking for the presence of countermeasures without attempting to gain unauthorized access to user accounts. For unchecked token authenticity, we closely monitored the experiments to prevent any unwanted side effects.

We did not compare our approach with a baseline because, to the best of our knowledge, there exist no available technique/tool in the literature that applies security testing for broken authentication in a fully automated manner.

\subsection{Experiment Results}

\begin{table}[t]
\renewcommand\arraystretch{1.1} % 
\centering
\caption{Validation results.}
\label{tab:results-vulnerability-detection}
% \resizebox{\textwidth}{!}{
\resizebox{0.48\textwidth}{!}{
\begin{tabular}{l|cc|cc|cc}
 & \multicolumn{2}{c}{\textbf{Credential}} 
 & \multicolumn{2}{|c|}{\textbf{Password}} 
 & \multicolumn{2}{c}{\textbf{Unchecked}} \\  
 & \multicolumn{2}{c}{\textbf{Stuffing}} 
 & \multicolumn{2}{|c|}{\textbf{Brute Forcing}} 
 & \multicolumn{2}{c}{\textbf{Token Auth.}} \\ 
 \cline{2-7}
\textbf{API} & \textbf{TP} & \textbf{FP} & \textbf{TP} & \textbf{FP} & \textbf{TP} & \textbf{FP} \\ \hline \hline
Here, ID4I, 6 Dot, BRAINBI & 1.0 & 0.0  & 1.0 & 0.0  & - & -  \\
BeezUP & 0.0 & 0.0  & 0.0 & 0.0  & - & -  \\
\hline
Tradematic & 0.0 & 0.0 & 0.0 & 0.0 & 0.0 & 0.0  \\
\end{tabular}}
\vspace{-3mm}
\end{table}

Table~\ref{tab:results-vulnerability-detection} presents the experimental findings, showing that both the credential stuffing and password brute forcing security testing strategies exhibit promising vulnerability detection capabilities.
In particular, four out of the six APIs
in the benchmark have been reported as vulnerable to credential stuffing and password brute forcing, and all the reported vulnerabilities have been manually confirmed, leading to a precision of the vulnerability detection of 100\%. The strategies successfully identified all vulnerabilities with no false positives.\footnote{We could not measure false negatives as there was no ground truth.} 
We attribute this success to the relatively straightforward nature of detecting these vulnerabilities. Nevertheless, the results underscore the effectiveness of testing strategies implemented with \securest and uncovering vulnerabilities that were not known in real-world APIs.
We have reported the detected vulnerabilities to the respective API owners.
% and are waiting for their acknowledgment.

We had only one API (\ie Tradematic) that was suitable for evaluating the unchecked token authenticity testing strategy.
The remaining APIs were either commercial and required a subscription or lacked valid credentials necessary for this assessment.
%
% This because the remaining APIs either did not allow the registration of new users since they are private, which made it impossible to gather the valid credentials needed for this assessment, or they were commercial APIs that required a subscription.
\securest correctly reported no vulnerabilities in Tradematic, indicating that the strategy is not prone to false positives. However, results from a single API case study may not fully represent the general behavior of the strategy, so further validation is necessary.

\section{Conclusion}\label{sec:conclusion}
We presented \securest, an open-source automated testing tool to uncover broken authentication vulnerabilities in Web APIs. We conducted an empirical validation and found that \securest is effective. Notably, it uncovered four previously unknown vulnerabilities in four publicly accessible APIs.

\securest is publicly available to encourage future research on this critical topic, provide collaboration opportunities with industry, and support the development of more secure web APIs.
Link to GitHub page: \url{https://github.com/SeUniVr/AuthREST}

%
% In future work, we intend to expand the empirical validation of the three existing testing strategies by incorporating additional subject APIs to validate and strengthen our findings. 
%
% We plan to collaborate with industry partners who are interested in utilizing \securest in their test environment and examine their authentication mechanisms. Moreover, we plan to enrich \securest with novel abstract and/or concrete components. 
% An integral part of this extension involves introducing a component capable of dynamically modifying the security testing strategies while HTTP interactions are ongoing, guided by real-time feedback from the observed interactions with the API under test. Lastly, our plans include the development of new automated security testing strategies that address emerging vulnerabilities, such as broken object-level authorization, broken object property-level authorization, and broken function-level authorization.

\bibliographystyle{IEEEtran}
\bibliography{main}

\end{document}